# The Korrontea Data Modeling

Emmanuel Bouix
LIUPPA – IUT de Bayonne
Château Neuf – Place Paul Bert
64100 Bayonne, France
+33(0)559574326

ebouix@acm.org

Philippe Roose
LIUPPA – IUT de Bayonne
Place Paul Bert
64100 Bayonne, France
+33(0)559574348

roose@iutbayonne.univ-pau.fr

Marc Dalmau
LIUPPA – IUT de Bayonne
Château Neuf – Place Paul Bert
64100 Bayonne, France
+33(0)559574321

dalmau@ieee.org

## ABSTRACT

Needs of multimedia systems evolved due to the evolution of their architecture which is now distributed into heterogeneous contexts. A critical issue lies in the fact that they handle, process, and transmit multimedia data. This data integrates several properties which should be considered since it holds a considerable part of its semantics, for instance the lips synchronization in a video. In this paper, we focus on the definition of a model as a basic abstraction for describing and modeling media in multimedia systems by taking into account their properties. This model will be used in software architecture in order to handle data in efficient way. The provided model is an interesting solution for the integration of media into applications; we propose to consider and to handle them in a uniform way. This model is proposed with synchronization policies to ensure synchronous transport of media. Therefore, we use it in a component model that we develop for the design and deployment of distributed multimedia systems.

## Categories and Subject Descriptors

D.2.2 [**Software Engineering**]: Requirements/Specifications – *Methodologies, Tools*.

## 1. INTRODUCTION

With the development of ubiquitous computing, multimedia data is now available on devices like mobile phones, PDA, and laptops. The Internet is a notorious wide area network used to transport this data between these devices. These possibilities create new needs for the deployment of distributed multimedia applications.

Our objective is to provide a global method for designing and developing multimedia applications. This work is QoS (Quality of Service) driven because these applications impose stringent requirements that the network layer of the Internet does not consider. Indeed, the quality required by end-users and the one provided by runtime environments are not taken into account. Thus, using these applications in such environments is compromised due to their moving and non-predictable characteristics (e.g. network bandwidth, terminal characteristics, operating system functionalities but also handicaps and languages of end-users). We define a software architecture suited to these applications [11]. Entities that compose it can be supervised by a middleware introduced to manage QoS [10], [12]. As an example of such applications, we can quote remote video monitoring which allows to monitor events or physical phenomena by using sensors like in [1]. We can use this kind of applications to keep watch on car parks or critical sections of roads where risks of traffic jam are higher [2]. Another more common example is videoconferencing systems which allow the meeting of several persons physically located in different places [3].

This paper focus on the specifications of a data model that we use for this architecture [11] in order to handle data and media. This model is specified with synchronization policies used at runtime to keep synchronization properties of data. We call it "Korrontea" which means "the data flow" in the Basque language.

The rest of the paper is structured as follows: Section 2 provides some justification on the needs of such a model. Section 3 presents the Korrontea data model and its main characteristics. In this section, we present policies used to ensure synchronization in multimedia applications too. Section 4 describes briefly the Osagaia component model specified to develop these applications, the aim is to show the use of Korrontea model. Section 5 presents the prototypes used to validate our works. Section 6 describes the related works. Section 7 provides some conclusions and discusses future work.

## 2. WHY DO WE NEED A MODEL?

By the mean of this part, we try to motivate our approach in detailing what we believe to be the important issues in modeling multimedia data. On one hand, we detail the media. On the other hand, we describe a brief survey of the architecture defined in our previous works. Then, we will be able to highlight the applications requirements in order to handle and integrate media.

### 2.1 The Media

The term media has a rich set of connotations. Media are form of information content where the goal is to inform or entertain end-users or audience. Media are related to how information is conveyed and distributed. They exist in different forms and are very used in applications. Several research works are interested in the classification of multimedia applications [4], [5]. Some of them show the importance of data in such applications. In [5], the authors define multimedia applications as an information processing system which handles a combination of media like e.g. text, graphics, images, audio, video or control information. They classify such systems by means of three criteria. The first criterion expresses the number of media used in an application; the second introduces the concept of time and distinguishes time-dependent and time-independent media; and the last means that the different types of media remain independent but can be processed and presented together. Combining all three criteria, they consider that these applications must support the integrated processing of several media types with at least one-time dependent one. In our works, we are interested by the systems of the same kind.

Some media are constituted of a sequence of media elements, also called samples, which describe an information part represented by the media in an adequate coding format. Often, they exist under the form of data flows. A data flow is a structure which provides information concerning the physical organization of samples, for instance their physical ordering and placement. Obviously, this kind of property must be considered in applications. These media are called continuous; they are based on human sensory properties. A particular characteristic is that they integrate synchronization relations between samples of both a single and several media [6]:

- intra-media refer to time relations between samples of the same media;
- inter-media refer to time relations between samples of several media.

These relations must be considered in order to achieve a natural impression at rendering time. The properties of physical ordering and synchronization bring an important part of the media semantics. Some studies on human perception of media and synchronization [7], [8] prove this viewpoint. Video and audio are examples of this kind of media.

Other kinds of media exist under different forms where the time factor is not preponderant. They are composed of a set of indivisible data necessary to render the media correctly. This is an essential property of this kind of media. They are called discrete media. An image is a discrete media composed of a finite set of pixels. Text and graphics are other examples.

Sometimes, it is necessary to synchronize discrete and continuous media in an inter-media way. This is the case of a video composed of audio, images and subtitles.

## 2.2 Functional Specifications of Multimedia Applications

The multimedia applications are designed according to a top-down decomposition. The goal is to obtain an application composed of a set of functional roles. Thus, we dispose of applications divided hierarchically into smaller and more manageable parts. This is an interesting issue for systems which plan to dynamically manage QoS. Such decomposition is described by means of graphs oriented and polar noted $G(V, E_s, E_c)$. We called them functional graphs; they are based on conditional process graph described in [9]. More details about them are done in [10]. The set V represents the nodes of the graph where each of them represents a basic role or functionality of an application noted $R_{i-j}$. Concretely, roles are performed by either software or hardware components. $E_s$ and $E_c$ are the sets of simple and conditional edges ($E_s \cap E_c = \emptyset$). Whatever its type, an edge $e_{ij}$ is used to link the output of node $P_i$ to the input of node $P_j$. An edge $e_{ij} \in E_c$ is a conditional edge where each of them has an associated condition value. The paths described by these edges can be considered only if the associated condition value is true. They allow to indicate different configuration choices and thus constraints in the using of particular roles. An edge $e_{ij} \in E_s$ is a simple edge. It means that the next node is an element of the application whatever the configuration. Edges represent media. They are used to connect the roles, i.e. components: this is horizontal composition.

The Figure 1 summarizes this approach by giving the representations used on functional graphs. The top right side shows the specification of a basic functionality whose function is to convert a color video media into a black and white one. On the bottom left side, we give an example of conditional edges (represented with thick lines) where three configurations are specified. The choice of one compared to another is specified by the conditions values noted $C_{F1}$, $C_{F2}$ and $C_{F3}$. Each of them is exclusive. According to both required and provided QoS, the platform will choose the configuration described by the path $C_{F1}$, $C_{F2}$ or $C_{F3}$. For instance, if the video must be rendered on a device with a restricted display capacity, we should choose the path corresponding to $C_{F3}$ which implements an image size reduction processing. The functional graphs allow to specify inter-media synchronization by means of synchronization links between edges. Such a link specifies the fact that media must be kept synchronous during their transport in applications. An example of such a specification is presented in the bottom right side of Figure 1 where audio and video must be kept synchronous in spite of the processing applied on video. Indeed, media processing introduces a problem that we identify as inter-media desynchronization. It is due to the fact that some media synchronized with others must be processed. Processing introduces temporal delays on media and so desynchronizes the processed media and the others. In the example of Figure 1, the video will be delayed by the image size reduction processing.

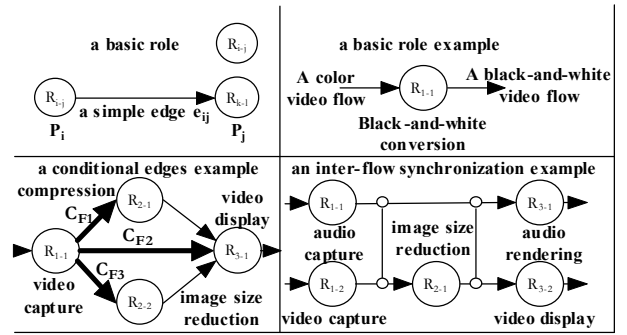

**Figure 1. Representation of Functional Graphs**

Another source of inter-media desynchronization can occur when synchronous media are transmitted through Internet network. Some services used to transport data on a network introduce an increase of the network load but also packet loss, delay and jitter [13]. These problems are well-known and harmful to media synchronization [14]. For example, the congestion control of TCP [15] increases transmission delays when errors were detected because of the retransmission mechanisms defined by this protocol. This introduces jitter in the media and so synchronization relations were altered.

## 2.3 Summary of Requirements

This study proves that the handling of media in applications is not an easy task. Media is a central concept. This data has properties that we should consider in order to avoid the loss of their semantics. Thus, we must integrate media at design time by means of the Korrontea model that we present in this paper. In Table 1, we summarize the requirements imposed by the handling of media and by the pattern given by the functional graphs.

**Table 1. Requirements and Suggested Solutions**

| Requirements/Specifications | Suggested Solutions |
|---|---|
| A structure which allows an easy integration of all the kinds of media | Media are existing in the form of data flows |
| Considering the physical ordering of data in a flow | Using an approach based upon a logical clock |
| Considering the intra-media synchronization relations | Using an approach based upon a physical clock |
| Allowing the handling and processing of media in applications | Defining the internal structure of data flows handled by implementation units |
| Allowing transmission of media through Internet network | Defining objects and mechanisms for this task |
| Classification of media | Based upon their temporal constraints |
| Ensuring inter-media synchronization | Defining Synchronization policies |

We propose a data model which meets these requirements. The aim is to define a structure which allows an easy integration of the media and data that may exist in distributed multimedia applications: we propose to use data flows. An advantage is to allow the integration of heterogeneous and interleaving media. Thus, all the data has sequence and synchronization relations properties. The data which composes data flows is ordered with a sequence number given by a logical clock. Moreover, time stamps are used in order to explicitly define and consider the synchronization relations. We define synchronization policies with the objective to keep synchronization between several media. Thus, definition of inter-media relations between several types of data becomes possible. These policies are based upon temporal behavior of data flows, i.e. the temporal constraints they integrate. By means of this model, we propose one way to handle process and transfer media into applications in both local and distributed cases.

In previous section, we identify two sources of desynchronization that we need to avoid. The policies introduced by Korrontea actually solve the problem. We detail the Korrontea model and its policies in the next section.

## 3. THE KORRONTEA DATA MODEL

We begin by describing the structure used to define data in multimedia applications. Before beginning details of this model, we give a representation of it under a formalism based on a UML [24] class diagram described in Figure 2. This diagram shows the data flow structure that we propose to use. The aim of this diagram is to illustrate the following definitions and properties.

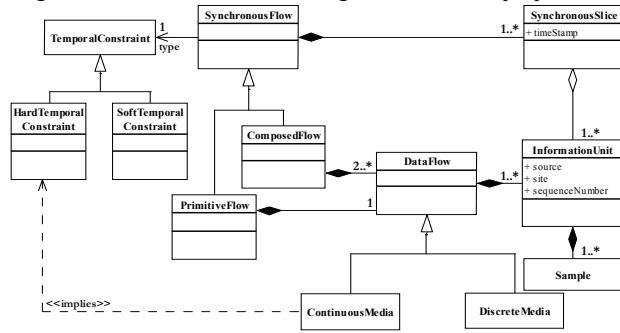

**Figure 2. The Korrontea Data Flows Conceptual Model**

### 3.1 A Common Structure: the Data Flows

The two kinds of media identified previously differ mainly by time constraints and data structure. This implies different ways for handling them and the necessity to know their types a priori. A lot of frameworks defined to provide multimedia programming proceed suchlike (see e.g. the sun one JMF [23]). Moreover, it is possible to define inter-media relations between data of different types. For facilitating this possibility and the handling of all kinds of media and data without a priori knowing their characteristics, we propose to use a unique abstraction which allows this in a uniform way: data flows. Such an abstraction is an interesting solution for the integration of media into these applications [5]. In functional graphs, we detail them by means of edges. Each edge takes its origin in one particular node of capture or creation. A data flow is produced by a unique component located in a unique site. These particular components are called located sources.

**Definition 1** *Located Sources*
*We call a located source LS a couple (S, L) where S is a component and L the site where S is located.*

An important notion brought by this component is the characteristic of distribution. Thus, these components are concrete entities of creation of data flows. An important property of these components is the site where they are located. We will see that this model is based onto it.

Data flows are composed of a sequence of data called samples. They include such things as video-frames, images, text, audio samples, events, etc.

**Definition 2** *Data Flows*
*A data flow f is composed of a possibly infinite sequence of samples with finite size. Each sample of f is produced by the same located source LS=(S, L).*
*Attributes of a data flow:*
- *locatedSource(f)=LS*
- *source(f)=S*
- *site(f)=L (capture or creation site of the flow)*

The creation of data flows by located sources consists in the continuous production of samples at either regular or irregular rate. Samples are produced in an adequate coding format which describes the information transported by the flow. We consider that all data flows are a priori composed of an infinite sequence of samples. Undoubtedly, data flows composed of a finite sequence can exist too, as in the case of data flows stored in files.

When data flows have no more samples that mean that this data is not used any more by an application. In such a case, these flows and components which produced and processed them are removed from an application. It is important to notice that the operators, properties and policies that we define can be only applied onto data flows which have samples.

At creation time but also during the handling and processing of samples (by the components of an application), the property of sequence must be explicitly considered. For that reason, we use a logical clock to stamp each sample or set of samples with an integer which allows to distinguish two successive samples or set of samples and so allows to maintain this property. Each component of an application provides, for each flow it produces, a mechanism which produces incremented integers. When processing samples, a counter "ticks": I=I+1. These integers are called sequence numbers. This concept was initially defined by Lamport [16] and classically used in distributed systems to preserve the order of data production [17], [18].

Handling data flows in applications depends on intrinsic characteristics of the supported information. A media can be seen according to several hierarchical levels where each of them describes a different granularity. For instance, a video can be decomposed into scenes, images, blocks and pixels. Applications are defined to handle one of these units. However, when a fine-grained one is chosen, like pixel for a video, it will be difficult or even impossible to keep the intra-media relations as software. Some works try to provide hardware solutions for the handling and processing of fine-grained units of media e.g. the works described in [19]. In order to avoid this problem, it is necessary to consider a sufficient number of samples. We propose to gather sufficient set of samples (sufficient grain) into information units. For instance, audio data are handled into programs by means of audio segment which gather samples which correspond to approximately 200 milliseconds [29]. This quantity of information is associated with the sequence number. In [5], the authors use nearly the same idea through the concept of Logical Data Units (LDU).

**Definition 3** *Information Units*
*We call an information unit IU, a couple (E, i) where E={$e_1$, $e_2$, ..., $e_n$} is a finite set of samples of the same data flow f and i an integer used to define the sequence property for this data flow. We will see that this property is defined between information units of the same data flow.*

*Attributes of information units:*
- *flow(IU)=f (data flow from whom elements of E belongs to)*
- *samples(IU)=E*
- *sequenceNumber(IU)=i*
- *locatedSource(IU)=locatedSource(flow(IU))*
- *source(IU)=source(flow(IU))*
- *site(IU)=site(flow(IU)) (capture or creation site of the unit)*

Thus, the constitution of information units depends on the considered media and on the specifications of the application. A sufficient number of samples is necessary. The size of information units depends on these specifications but also on characteristics of data. We propose to put in information units all the samples created by a located source at the same time. For example, in a video we gather all the pixels which constitute an image. According to the specifications, we can decide too to compose information units by one, two or three images, etc.

When transporting data flows into applications, information units are read and written in isolation or in synchronous sequences and this in a continuous way. Processing and transport will be triggered when information units are available in the entities in charge of these tasks. We use a software architecture similar than the pipes & filters one [25].

Therefore, data flows are composed of sequence of information units. This kind of structure is linked to the concept of time. Two successive units are separated by certain durations. The creation of a unit is done at a particular time earlier than the creation of the next one. This is the intra-media synchronization relations defined previously. By using such a structure, these relations are now properties of the data flows. Physical clocks are mechanisms used to define temporal values and to stamp data in distributed systems. The main difference with logical one is that with a physical clock, we manipulate real time. Real time is a continuous dense set not limited of time values which represent physical instants. Dense set means that there is always at least one instant between any pair of instants. Thus, we can handle the concepts of time value, time interval and duration. More details about physical clocks can be found in [20] where a time model is defined. Physical clocks can be global or local. Global ones allow to dispose of a unique temporal reference in distributed systems. A time value provided by these clocks describes the same physical instant on every site where such a system is deployed. Several research works deal with this solution. Their definition is based upon clocks at each site. These approaches impose approximations due to temporal skews of clocks in relation with others [21]. This research is based upon both probability and statistical approaches. Such a solution will probably influence the degree of synchronization because it is impossible to have a perfect and absolute synchronization of clocks through the Internet network [22]. Consequently, it is difficult to use a global clock without introducing a margin of error. A second approach consists in using local physical clocks. This solution recommends the use of physical clocks at each site where a system is deployed. The same concepts can be handled but in this case relatively to the clock of a particular site: temporal relations can be defined and kept between data created on the same site. We consider that the synchronization of media created or captured on different sites is something artificial and so it is not necessary to have strict temporal relations for this task. Our purpose is not to create synchronization but to keep it in order to prevent the loss of information (the one of synchronization) which existed at creation time of several media coming from the same site. Indeed, there are not many cases where artificial synchronization must be performed. Moreover, the component model that we define allows to create artificial synchronization by using processing of media (see the application example in the following paper [12]). On the basis of these arguments, we choose to use an approach based upon local physical clocks. Thus, located sources presented in definition 1 allow to introduce this notion of localization. These particular components take their meaning in the following definitions and properties.

**Definition 4** *Local Physical Clock*
*We assume that in every site L, there is a unique component $C_p$ (local physical clock of L). Samples produced by LS=($C_p$, L) are integers strictly increasing called time stamps.*

These integers considered independently of each other have no meanings. Nevertheless, they are linked to real time by a proportionality ratio. Thus, if LS produces integer $n_1$ at the time value $t_1$ and integer $n_2$ at the time value $t_2$, then $(n_2-n_1)=k_L(t_2-t_1)$ where $k_L$ is a constant known in site L.

**Assumption 1** *Time Constant*
*On each site of a multimedia application, constant $k_L$ is the same and is equal to k.*

This assumption means that all the local physical clocks of an application have the same rate. Time values produced by local physical clocks ($C_{p1}$, $L_1$) and ($C_{p2}$, $L_2$) cannot be compared; we can only compare differences between time values produced respectively by these located sources. This is an important characteristic of our model. Indeed, knowing the constant $k_L$ allows to respect intra-flow synchronization relations at rendering time and whatever the site where data flows come from.

Synchronization between data flows which come from the same site can be kept during their transport with the time stamps done by local physical clocks. We define an object that we call synchronous slice. It is used to gather a set of information units created or captured in the same site and which corresponds to the same time interval.

**Definition 5** *Synchronous Slices*

Let $F=\{f_1, f_2, ..., f_n\}$ a set of data flows such as $\forall f_i \in F$, $site(f_i)=L$. We call a synchronous slice, defined on F, an object which contains an information quantity which corresponds to the same time interval and that we define as follows: $SS=(t, U)$ where t is a time stamp created by $LS=(C_p, L)$ and U is a finite set of information units such as $\forall x \in U$, $flow(x) \in F$. Sequence numbers of information units are defined in order to respect the following property: let $A_f=\{e \in U\ /\ flow(e)=f\}$ then $\forall e \in A_f$, $1 \leq sequenceNumber(e) \leq |A_f|$ (cardinal of set $A_f$) and $\forall e_1, \forall e_2 \in A_f$, $e_1 \neq e_2 \Leftrightarrow sequenceNumber(e_1) \neq sequenceNumber(e_2)$.

*Attributes of Synchronous Slices:*
- $flow(SS)=F$ (this attribute allows to know the data flows of a particular synchronous slice)
- $timeStamp(SS)=t$
- $informationUnits(SS)=U$
- $site(SS)=L$

The time stamps are assigned to slices after the creation of information units that will be gathered in them. In a same slice, the information units are stamped with sequence numbers. The sequence number is equal to one for the first information unit and is equal to n for the last information unit by considering that the slice contains n information units.

Synchronous slices are the units of synchronous transport of data in applications. Application components exchange synchronous slices between one another. Then, components retrieve the information units of flows they will process. Others are in slices only to keep synchronization relations. We define synchronous flows that are composed of a sequence of synchronous slices.

**Definition 6** *Synchronous Flows*
A synchronous flow SF is composed of a possibly infinite sequence of synchronous slices with the same flow set such as $\forall SS \in SF$, $flow(SS)=F$ (see definition 5). Synchronous slices of a same synchronous flow have different time stamps, i.e.: $\forall SS_1, \forall SS_2 \in SF$, $SS_1 \neq SS_2 \Leftrightarrow timeStamp(SS_1) \neq timeStamp(SS_2)$.

*Property of Synchronous Flows:*
- $\forall SS_1, \forall SS_2 \in SF$, $site(SS_1)=site(SS_2)=L$
- *Proof:* According to this definition, we have $\forall SS_1, \forall SS_2 \in SF$: $flow(SS_1)=flow(SS_2)=F$. According to definition 5, F is a set of data flows such as $\forall f_i \in F$, $site(f_i)=L$

*Attributes of Synchronous Flows:*
- $flow(SF)=F$ where F is the set of data flows (see definition 2) which composes synchronous flows
- $site(SF)=L$ (all the flows, which compose synchronous slices of SF, come from the same site)

An information unit is an element of one and only one synchronous slice. In the same way, a synchronous slice is an element of one and only one synchronous flow.

**Property 1** *Property of Information Units Membership*
- $synchronousSlice(IU)=SS \in SF$ such as $IU \in informationUnits(SS)$

According to the composition of synchronous slices, we distinguish two kinds of synchronous flows: the primitive one and the composed one.

**Definition 7** *Primitive Synchronous Flows*
A synchronous flow is defined as primitive when $|flow(SS)|=1$.

Data flows linked by inter-flow synchronization relations are put in composed flows. For each data flow, synchronous slices of composed flows contain information units which correspond to the same time interval. The provided policies that we will describe are used to create composed flows.

**Definition 8** *Composed Synchronous Flows*
A synchronous flow is called composed when $|flow(SS)|>1$.

All data will exist under the form of synchronous flows.

Information units of synchronous flows can be ordered by time and by an integer which represents the sequence. We define strict total order relations between information units of a same synchronous flow. This property can be used at presentation time of data in order to have ordered rendering.

**Property 2** *Strict Total Order Relations (< and >) between information units of a data flow into a synchronous flow SF*
We define < (respectively >) as a strict total order relation between information units of a same synchronous flow. Let $IU_1$ and $IU_2$ two information units $\in SF$, with $flow(IU_1)=flow(IU_2)$. We have:
- $t_1=timeStamp(synchronousSlice(IU_1))$, $n_1=sequenceNumber(IU_1)$
- $t_2=timeStamp(synchronousSlice(IU_2))$, $n_2=sequenceNumber(IU_2)$

$IU_1<IU_2$ (respectively $IU_1>IU_2$) $\Leftrightarrow$ $(t_1<t_2)$ OR $((t_1=t_2)$ AND $(n_1<n_2))$ (respectively $(t_1>t_2)$ OR $((t_1=t_2)$ AND $(n_1>n_2)))$. If $t_1 \neq t_2$, the order is done by the time stamp. If $t_1=t_2$, the order is done by the sequence number.

*Proof:* A strict order has properties of irreflexivity and transitivity. Moreover, this order is total if we have: $\forall IU_1$ and $IU_2$ two information units $\in SF$, with $flow(IU_1)=flow(IU_2)$, $IU_1<IU_2$ OR $IU_1=IU_2$ OR $IU_2<IU_1$.

- *Irreflexivity:* $\forall x \in SF$, let SS a synchronous slice / $x \in informationUnits(SS)$.
  $timeStamp(synchronousSlice(x))<timeStamp(synchronousSlice(x))$ is false because these two time stamps are equals. In the same way, $sequenceNumber(x)<sequenceNumber(x)$ is false because these two numbers are equals. Hence, relation $x<x$ is false.
- *Transitivity:* $\forall IU_1$, $IU_2$ and $IU_3$ three information units $\in SF$, with $flow(IU_1)=flow(IU_2)=flow(IU_3)$ such as $IU_1<IU_2$ and $IU_2<IU_3$. We have:
  $t_1=timeStamp(synchronousSlice(IU_1))$, $n_1=sequenceNumber(IU_1)$ and
  $t_2=timeStamp(synchronousSlice(IU_2))$, $n_2=sequenceNumber(IU_2)$ and
  $t_3=timeStamp(synchronousSlice(IU_3))$, $n_3=sequenceNumber(IU_3)$.
  - if $IU_1<IU_2$ is due to the fact that $t_1<t_2$, $IU_2<IU_3 \Rightarrow t_2 \leq t_3$, so we have $t_1 < t_2 \leq t_3 \Rightarrow IU_1<IU_3$.
  - if $IU_1<IU_2$ is due to the fact that $t_1=t_2$ AND $n_1<n_2$ and $IU_2<IU_3$ is due to the fact that $t_2<t_3$, so we have $t_1=t_2<t_3 \Rightarrow IU_1<IU_3$. If $IU_2<IU_3$ is due to the fact that $t_2=t_3$ AND $n_2<n_3$, so we have $t_1=t_2=t_3$ and $n_1<n_2<n_3 \Rightarrow IU_1<IU_3$

Hence, < is a strict order relation. The same proof can be established for relation >.

Now, we must prove that these relations are total. Let $t_1=timeStamp(synchronousSlice(IU_1))$, $n_1=sequenceNumber(IU_1)$ and $t_2=timeStamp(synchronousSlice(IU_2))$, $n_2=sequenceNumber(IU_2)$. $t_1$ and $t_2$ are produced by a physical clock, they are integers and we know that < is a total order relation on integers, so we have:
- $t_1<t_2 \Rightarrow IU_1<IU_2$
- or $t_2<t_1 \Rightarrow IU_2<IU_1$

- or $t_1=t_2$: $IU_1$ and $IU_2$ are information units which verify $flow(IU_1)=flow(IU_2)=f$. According to the definition 5, we know that $\forall\ e_1,\ \forall\ e_2\ \in\ A_f,\ e_1 \neq e_2 \Leftrightarrow sequenceNumber(e_1) \neq sequenceNumber(e_2)$, so we know that $IU_1 \neq IU_2 \Leftrightarrow n_1 \neq n_2$ and by contraposition $IU_1=IU_2 \Leftrightarrow n_1=n_2$. The relation < is a total order on integers, so we have:
  - $n_1<n_2 \Rightarrow IU_1<IU_2$
  - or $n_2<n_1 \Rightarrow IU_2<IU_1$
  - or $n_2=n_1 \Rightarrow IU_2=IU_1$

Thus, < (respectively >) is a strict total order relation.

The relation < means "earlier than" and the relation > means "later than". In the same way, we define strict total order relations between synchronous slices of synchronous flows.

**Property 3** *Strict Total Order Relations (< and >) between synchronous slices of a synchronous flow SF*

*We define < (respectively >) as a strict total order relation between synchronous slices of a same synchronous flow by:* $\forall\ SS_1 \in SF,\ \forall\ SS_2 \in SF,\ SS_1<SS_2$ *(respectively* $SS_1>SS_2$*)* $\Leftrightarrow timeStamp(SS_1) < timeStamp(SS_2)$ *(respectively* $timeStamp(SS_1) > timeStamp(SS_2)$*).*

<u>Proof:</u> $\forall\ SS_1,\ SS_2 \in SF$ with $SS_1 \neq SS_2$, we have $timeStamp(SS_1) \neq timeStamp(SS_2)$ (see definition 6). Moreover, timeStamp() is an integer and so we have $timeStamp(SS_1) < timeStamp(SS_2)$ or $timeStamp(SS_1) > timeStamp(SS_2)$. We have also $SS_1<SS_2$ or $SS_1>SS_2$. Thus, all synchronous slices of a synchronous flow can be ordered by relations < and > $\Rightarrow$ these relations are strict total order between synchronous slices of a synchronous flow.

These relations are defined on objects that compose synchronous flows.

**Property 4** *Properties of synchronous flows (part 2)*

*Synchronous slices of a synchronous flow are totally ordered by < and > strict total order relations (see property 3). Information units of slices are totally ordered by < and > strict total order relations (see property 2).*

Thanks to these relations, we can handle sequence and time. We define operators which permit, for each synchronous slice, to handle the sequence in flows. For each slice we can know the previous and the next ones.

**Definition 9** *Previous (prev) and Next (next) Operators*

*According to the property 3, synchronous slices of a synchronous flow SF are totally ordered by < and > relations. So, we can define:*

- $\forall\ SS_1 \in SF,\ \exists\ SS_2 \in SF$ called $prev(SS_1)\ /\ SS_2<SS_1$ and $\forall\ SS_3 \in SF$, we have: $SS_3<SS_2$ or $SS_3>SS_1$
- $\forall\ SS_1 \in SF,\ \exists\ SS_4 \in SF$ called $next(SS_1)\ /\ SS_4>SS_1$ and $\forall\ SS_3 \in SF$, we have: $SS_3>SS_4$ or $SS_3<SS_1$

We can handle time intervals too. The using of a local physical clock imposes that these intervals are defined between slices with the same site of capture/creation.

**Definition 10** *Time Intervals*

*We define time intervals between synchronous slices of a synchronous flow SF as follows:* $\forall SS_1, SS_2 \in SF$, *we define* $<SS_1, SS_2>=|timeStamp(SS_2)-timeStamp(SS_1)|$.

The physical clock introduces explicit temporal behavior in synchronous flows. Indeed, all these flows have different intra-flows synchronization relations. Thus, we propose to distinguish the synchronous flows according to their temporal constraints. These constraints are detailed in the next section.

## 3.2 The Temporal Constraints

Previously, we described continuous and discrete media. This distinction shows the diversity of media and specific characteristics of each of them. However, we think that this classification is not perfect. For instance, a media composed of sub-titles of a video describes the characteristics of both continuous and discrete media. The rendering of sub-titles must respect precise time values and this kind of media is sensitive to data loss. In the same way, a slide presentation is seen as an image flow with intra-flow relations less strict than in a video but the loss of an image may damage the semantics of the presentation.

The last section introduces a unique structure for these media. According to the kinds of data and media supported, the temporal behavior of flows will be different. The classification of media into continuous and discrete categories is not relevant to our works. A slide presentation where images have intra-flow relations greater than 10 seconds is considered as a discrete media. What happens if in the same kind of flow, the intra-flow relations are lower than the second? Consequently, we do not consider this criterion. We do not keep the criterion of the regularity of data into media because this solution considers in the same way an image flow where images are separated with regular rate of one minute and an image flow where images are separated at a regular rate of one second. Even if these two examples of flow are regular, they are different to the extent that they integrate different temporal behaviors.

Thus, it appears that media can be classified according to their temporal behaviors. Undoubtedly, this way of thinking is not universal and so cannot be defined strictly but preferably at specification time. This implies that such a classification depends on the specifications of applications to develop. We mean it depends on the media used. The relevance of this classification is guided by the fact that the temporal behavior defines the way of handling media. Thus, we distinguish synchronous flows with either soft or hard temporal constraints. We consider that continuous media have hard temporal constraint due to their types of information and to their types of temporal constraints (studies on human perception of media argue this viewpoint [7], [8]) and also because non-observance of these constraints involves the loss of the media semantics. In the UML diagram, we specified this property by means of an <<implies>> dependency [24] (see Figure 2). On the other hand, discrete media can be with hard or soft temporal constraints.

**Property 5** *Hard Temporal Constraint*

*A synchronous flow has hard temporal constraints if, $\theta$ being defined for an application,* $\forall SS \in SF\ /\ <SS, next(SS)> \leq \theta$.

Previously, we have seen that some kinds of discrete media can have this temporal constraint.

For distinguishing the temporal constraints of data, we introduce a parameter $\theta$ which must be defined at design time. It represents the maximal time value between two successive synchronous slices. It depends on the media used in applications. Thus, we cannot give a particular value. When the value between two successive slices is greater than $\theta$ then the flow is considered with soft temporal constraint.

**Property 6** *Soft Temporal Constraint*

*A synchronous flow has soft temporal constraints if, $\theta$ being defined for an application,* $\exists SS \in SF\ /\ <SS, next(SS)> > \theta$.

This classification is used for handling media. The synchronization policies that we will define in the next section are based upon this. Moreover, components of an application will know how to handle data thanks to these constraints.

## 3.3 Multimedia Synchronization Policies

We saw that it is essential to ensure synchronization. Indeed, without synchronization the media look somewhat artificial and incomprehensible [7], [8]. Functional graphs allow specifying inter-flow relations. This specification means that flows linked by this way must be kept synchronous during their transport into applications. Intra-flow relations are not clarified because they are explicitly defined in synchronous flows by means of time stamps. These policies permit to keep synchronization relations during the transport of flows despite processing and network transfer. Synchronization relations are important at media rendering time. We detail the way that we use for keeping these relations.

### 3.3.1 Intra-Flow Synchronization

These relations correspond to the rate of flows; they give temporal relations between data which compose a flow. For instance, a 25 image per second rate video needs displaying one image every 40 milliseconds. These relations are not strict, tolerances may be accepted [7], [8]. This model gives characteristics to ensure these relations at rendering time. Thanks to order properties, information units can be rendered in an ordered sequence. The rendering of synchronous flows must be defined at design time and implemented in rendering components. For each synchronous flow, we can define intra-flow synchronization by using the sequence number of information units and the time stamp of synchronous slices. The strict total order relations permit to order all information units of these flows and so all samples. Temporal constraints of these flows determine the kind of intra-flow synchronization in relation with a time value $\theta$ defined at design time (see properties 5 and 6).

Sequence numbers, time stamps and temporal constraints are defined for each flow at creation or capture time by an adequate component called a located source (see definition 1). Some processing and handling of flows can affect this information. In order to avoid this problem, we are planning to update, by particular services attached to the components, these characteristics during the transport of flows into applications. These services are described by the component model that we define for the implementation of these applications [12]. This information must be used in an efficient way by all components that implement an application.

On the one hand, it is not a hard task to retrieve this kind of relations at rendering time. Indeed, the model is designed in this way. On the other hand, it is difficult or even impossible to retrieve the inter-flow ones if we do not provide mechanisms in order to keep them. We provide policies for this special purpose.

### 3.3.2 Inter-Flow Synchronization

Inter-flows synchronization corresponds to temporal relations that may exist between data of several flows. This is for instance the relations that link audio and images in a video. They are described in functional graphs by mean of synchronization links (see Figure 1). In our works, this kind of synchronization can be ensured between flows captured or created in the same site because our purpose is not to create inter-flow synchronization but to maintain it. Indeed, on creation or capture site, flows may have temporal and semantic relations.

Keeping these relations requires expressing relations between synchronous slices of different flows. We define these relations by providing operators that return particular time stamps. Then, these time stamps are used by policies in order to use synchronous slices which correspond to these particular time stamps. With this aim, we define minimalTimeStamp and maximalTimeStamp operators.

**Definition 11** *minimalTimeStamp and maximalTimeStamp Operators*
*We define operators that return the synchronous slices with the smallest and the greatest time stamp from a set of slices E. $E=\{e_1, e_2, ..., e_n\}$ with $\forall e_i, \forall e_j \in E$, $site(e_i)=site(e_j)$, such as:*

- $minimalTimeStamp(E) = min_{e_i \in E}(timeStamp(e_i))$

- $maximalTimeStamp(E) = max_{e_i \in E}(timeStamp(e_i))$

Several synchronous slices issue from the same site can have the same time stamp. Consequently, more than one element of a set E can match to minimalTimeStamp(E) (respectively maximalTimeStamp(E)).

These operators are defined to be applied on a set of synchronous slices. Synchronous flows are composed of set of synchronous slices. Thus, these operators can be applied on one or several synchronous flows. We define an operator which allows to obtain on such a set the minimal time stamp.

**Definition 12** *First Occurrence Operator on a set of Synchronous Flows*
*Let $SG=\{f_1, f_2, ..., f_n\}$ a set of synchronous flows such as $\forall f_i \in SG$, $site(f_i)=L$. On such a set, we define an operator called $firstOccurence_{SG}(t)$ as follows:*

- $firstOccurence_{SG}(t)=minimalTimeStamp(E)$ with $E=\{e_i \in f \mid timeStamp(e_i) \geq t \text{ with } f \in SG\}$

In the same way, we define an operator which allows to obtain on such a set the maximal time stamp.

**Definition 13** *Last Occurrence Operator on a set of Synchronous Flows*
*Let $SG=\{f_1, f_2, ..., f_n\}$ a set of synchronous flows such as $\forall f_i \in SG$, $site(f_i)=L$. On such a set, we define an operator called $lastOccurence_{SG}(t)$ as follows:*

- $\forall f \in SG$, we define the set $firstSlice_f(t)$ by: $firstSlice_f(t)=\{SS \mid SS \in f \text{ such as } timeStamp(SS) \geq t \text{ and } timestamp(prev(SS)) < t\}$. This set contains first synchronous slice of f which time stamp is greater or equal to t.
- $lastOccurence_{SG}(t)=maximalTimeStamp(E')$ with $E' = \bigcup_{f_i \in SG} firstSlice_{f_i}(t)$

We give an example of these two operators in Figure 3. This figure shows a set of synchronous flows. We show the slices whose time stamps correspond to first occurrence and last occurrence: respectively $i_1$ and $h_1$.

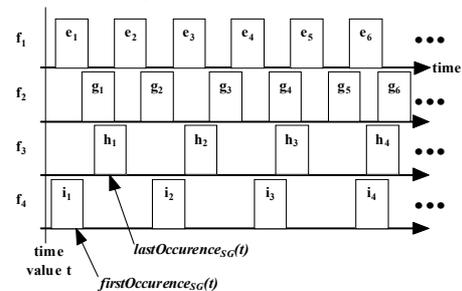

**Figure 3. First and Last Occurrences on a set of Flows**

These operators have several properties that we define as follows.

**Property 7** *Properties of Occurrence Operators*
- $lastOccurence_{SG}(t) \geq firstOccurence_{SG}(t) \geq t$
- <u>Proof:</u> $\exists\ e \in f$ such as: $timeStamp(e) = minimalTimeStamp(E) = \min_{e_i \in E}(timeStamp(e_i))$ We will demonstrate that $e \in firstSlice_f(t)$. We know that $timeStamp(e) \geq t$ (see definition 13). $prev(e)$ verifies property $timeStamp(prev(e)) < t$ because if not that means that $prev(e) \in E$ and in this case we would have $minimalTimeStamp(E)=timestamp(prev(e)) < timeStamp(e)$ because $prev(e)<e$ (see definition 9). Hence, we have $e \in firstSlice_f(t)$, then $lastOccurrence_{SG}(t)=maxsimalTimeStamp(E')$ with $E' = \bigcup_{f_i \in SG} firstSlice_{f_i}(t)$, so $e \in E'$ and $maximalTimeStamp(E') \geq timeStamp(e)$. Hence, we have $lastOccurence_{SG}(t) \geq firstOccurence_{SG}(t)$. Moreover, $firstOccurrence_{SG}(t)=minimalTimeStamp(\{e_i \in f\ /\ timeStamp(e_i) \geq t\}) \geq t$ because it is defined such as synchronous slices which have the minimum time stamp greater or equal to t.

The aim of these policies is to define composed flows and so their synchronous slices. Composition depends on the temporal constraints of the synchronous flows used. Policies are applied on set of synchronous flows. Such a set can contain indifferently primitive or composed flows, they produce composed flows. The following property will help to distinguish the different policies that we introduce.

**Property 8** *Property of a set of Synchronous Flows*
- A set of Synchronous Flows can be decomposed into two subsets as follows: $SG=\{SG_{hard}\} \cup \{SG_{soft}\}$. Thus, $SG_{hard}=\{f_i \in SG\ /\ f_i$ is a flow with hard temporal constraint$\}$ and $SG_{soft}=\{f_i \in SG\ /\ f_i$ is a flow with soft temporal constraint$\}$.

This property is important because it allows to define the way to constitute the synchronous slices of the composed flows. Starting from a set of synchronous flows, policies allow to make up composed flows. According to the constraints of synchronous flows, we define three different policies:
- the first one is applied when the set of synchronous flows contains only flows with hard temporal constraint;
- the second one is applied when the set of synchronous flows contains flows with both constraints;
- the third one is applied when the set of synchronous flows contains only flows with soft temporal constraint.

An advantage of our policies is that we link them with jitter compensation mechanisms. Indeed, we consider the fact that synchronous flows can arrive ahead of time or after the ones in relation with the others. We take into account these possible temporal delays between flows and so propose efficient policies. To do this, we introduce a maximum delay $\alpha$ which represents the maximum delay of synchronous slices on each flow. After, this delay we consider that in all the flows of the set all slices are available. We will see in each policy how to use this maximum delay. Each policy is detailed with its corresponding algorithm.

### 3.3.2.1 Hard Policy: $SG_{hard} \neq \varnothing$ and $SG_{soft}=\varnothing$

The principle of hard policy is to make synchronous slices of the resulting composed flow by gathering every slice of whose time stamp is included between the first occurrence and the last occurrence. This guarantees that every resulting slice will contain at least one information unit of each flow (see definition 13). In synchronous flows with hard temporal constraint, we are sure to receive a slice with a maximal delay of $\alpha+\theta$ (see properties 5 and 6). Starting from an instant t, we can wait until composing the set $firstSlice_f(t)$ for each flow of the set of synchronous flows.

The hard policy algorithm is done in Figure 4. The first task consists in waiting at least one synchronous flow on each flow. On the set formed by these slices, we apply the operator lastOccurence in order to determine a time value called $T_{MAX}$. In order to compensate time delays between flows, we wait for the slices which verify $timeStamp(SS) > T_{MAX}$. Once these tasks have ended, we can constitute the resulting slice by adding every information unit of the received slices which verify $timeStamp(SS) \leq T_{MAX}$. Then, we assign successive sequence number to each information unit. The time stamp of the resulting slice will be equal to firstOccurence applied to the slices used to compose it.

```
repeat
  - wait until each flow ∈ SG has at least one synchronous slice (each flow has its set firstSlice_f(t) defined)
  - T_MAX ← lastOccurence({every slice received})
  - wait until each flow has a synchronous slice SS which verifies timeStamp(SS) > T_MAX
  - result slice is composed as follows:
      - by adding every information unit of the slices received which verify timeStamp(SS) ≤ T_MAX
      - for each information unit of each flow, we assign successive sequence numbers (initialized to 1 for each flow)
      - by adding a time stamp = firstOccurence({synchronous slices used})
end repeat
```

**Figure 4. Hard Policy Algorithm**

### 3.3.2.2 Mixed Policy: $SG_{hard} \neq \varnothing$ and $SG_{soft} \neq \varnothing$

In mixed policy, we compose resulting slices by gathering every slice of synchronous flows where the time stamp is included between firstOccurrence and lastOccurrence applied in the subset SGhard. This guarantees that the resulting slice contains at least one information unit on each flow element of SGhard (see definition 13). Nevertheless, flows element of SGsoft can have information units or not. In this policy, we use a delay equal to $\alpha$ in order to wait for synchronous slices of flows with soft temporal constraint, beyond which we consider that all the slices of these flows have been received.

This algorithm is more complex than the previous because it is necessary to consider the flows with soft temporal constraint where it is not possible to know a priori if the fact that data is available or not. Thus, we propose to use a principle similar to the one of semaphore in operating system field. Each arrival of a slice on one of the flow with hard temporal constraint triggers a delay $\alpha$. At the end of this delay, an authorization of slice constitution is done. This mechanism allows to take into consideration the possible delay of receiving slices of same time stamp in other flows. Every slices received during this delay will not be obligatorily put in the same slice than the one which triggered the delay. Indeed, they will put in the same slice only if their time stamps verify the adequate properties.

The algorithm of this policy is done onto Figure 5. First, we initialize the variable "autorisation" to 0. Then, we wait on each flow with hard temporal constraint at least one synchronous slice. When a slice is available on these synchronous flows, we run a timer with a delay $\alpha$. At each end of the delay, the process executes: autorisation ← autorisation + 1. $T_{MAX}$ is defined with the operator lastOccurence applied on every synchronous slices received. We wait until each flow with a hard temporal constraint has a synchronous slice SS which verifies $timeStamp(SS)>T_{MAX}$. Finally, the resulting slices are composed by adding every information unit of slices received whose time stamps verify timeStamp(SS) ≤ $T_{MAX}$. Each of these units is associated with a sequence number initialized to 1 for each flow. The time stamp of the resulting slice will be equal to firstOccurrence applied on synchronous slices of flows with hard temporal constraint used. The variable "autorisation" is decreased by the number of slices of flows with hard temporal constraint used.

```
- autorisation ← 0
repeat
  repeat
    - when a synchronous slice is available onto one of flows ∈ SG_hard, run a timer with a delay α
  until we dispose of firstSlice_f(t) for each flow ∈ SG_hard
  end repeat
  - T_MAX ← lastOccurrence({synchronous slices received})
  - wait until each flow ∈ SG_hard has a synchronous slice SS which verifies timeStamp(SS) > T_MAX
  - wait until autorisation > 0
  - result slice is composed as follows:
    - by adding every information unit of slices received (onto both SG_hard and SG_soft) which verify timeStamp(SS) ≤ T_MAX
    - for each information unit of each flow, we assign successive sequence numbers (initialized to 1 for each flow)
    - by adding a time stamp = firstOccurrence({synchronous slices of flows ∈ SG_hard used})
  - autorisation ← autorisation - number of slices of flows ∈ SG_hard used
end repeat
At each end of timer, the process executes: autorisation ← autorisation +1
```

**Figure 5. Mixed Policy Algorithm**

### 3.3.2.3 Soft Policy: $SG_{hard}=\varnothing$ and $SG_{soft}\neq\varnothing$

The last policy consists in composing the resulting slices by gathering every received slice whose time stamp is included between firstOccurrence and firstOccurrence+θ. This guarantees that a slice will contain at least one information unit (see definition 12). In this case, we wait for the first synchronous slice which allows defining firstOccurrence.

We put in the resulting slices every slice whose time stamp does not exceed firstOccurence of a delay θ. Using θ, which constitutes the limit between both hard and soft temporal constraints, implies that we consider synchronous two slices produced in this time interval.

In this policy, we use the same mechanism than in the previous one. The authorization of the slices constitution is now based on a delay α+θ. It corresponds to the time to wait chosen (θ) augmented by the maximum delay α.

The soft policy algorithm is detailed in Figure 6. We initialize the "autorisation" variable to 0. As soon as a slice is available on one of the synchronous flows with soft temporal constraint, a timer is run with a delay α+θ. When autorisation is strictly greater than 0, we can begin to compose the resulting slice by adding every slice received which verifies timeStamp(SS) ≤ firstOccurence({slices received})+θ. Like in other policies, information units are associated with a sequence number. The time stamp of the resulting slice is equal to firstOccurrence applied on slices used to compose it. Finally, the variable "autorisation" is decreased by the number of slices used.

```
- autorisation ← 0
repeat
  - when a synchronous slice is available onto one of flows ∈ SG_soft, run a timer with a delay α+θ
  if autorisation > 0
    - result slice is composed as follows:
      - by adding every information unit of slices received which verify timeStamp(SS) ≤ firstOccurence({synchronous slices received})+θ
      - for each information unit of each flow, we assign successive sequence numbers (initialized to 1 for each flow)
      - by adding a time stamp = firstOccurrence({synchronous slices used})
    - autorisation ← autorisation - number of slices of flows used
  end if
end repeat
At each end of timer, the process executes: autorisation ← autorisation +1
```

**Figure 6. Soft Policy Algorithm**

### 3.3.2.4 Notice

The mechanism used to manage temporal delays is based upon the semaphore mechanism. At each end of the delay, we give an authorization to make up a slice by executing autorisation ← autorisation + 1. In the same way, each composition of the resulting slices corresponds to the consumption of such authorizations.

The delay α introduced to compensate delay between flows allows to synchronize these flows by considering margin of errors. If we choose for α a low value, we do not wait all the slices on all the flows and so the temporal relations between flows will introduce delays. If we choose for α a great value, we wait for all the slices and so the synchronization will be efficient without margin of errors. However, a too great value for α will increase latency in applications. We must find a compromise in the choice of a value. We are developing a prototype in order to perform this study.

## 4. THE OSAGAIA COMPONENT MODEL

The last part of our work is to define a software component model in order to ensure the implementation of distributed multimedia applications in accordance with the specifications given by functional graphs. Both design and data models give some specifications to define component model. This model is called "Osagaia" which means "the software component" in the Basque language. The nodes of graphs will be implemented by software components. These components will be connected by means of connectors whose role is to transport the media according to the Korrontea model. More principles about this model are presented in [12].

We identify two kinds of software components: the functional ones and the nun-functional ones. Functional components are in charge of the implementation of the basic functionalities of an application (defined by nodes of functional graphs). Frequently, they are components of creation/capture, processing, rendering or storage. Non-functional components are in charge of the implementation of the aspects which associate non-functional properties to the applications necessary for their implementation [26]. Among these non-functional components, we define a component called "fusion" whose goal is to produce composed flows from a set of synchronous flows by applying the synchronization policies we defined previously. In fact, this component allows defining synchronization links introduced by functional graphs. A component which realizes the opposite function is also defined, it is called "separation". It is used to break composed flows into primitive ones. Other non-functional are used but not described here.

Functional components are called business component since they implement the business functionalities of multimedia applications. Each of these functionalities is implemented by means of a component (some examples of such multimedia functionalities can be found in [27]). These components need to be executed in a container whose role is to provide non-functional implementation for components. Thus, a container is used to run a business component. The architecture of this container is shown in Figure 7, we call it the elementary processor. Its role is to perform interactions between business components and their outside. It is divided into two main parts: the exchange unit (composed of input and output units, see Figure 7) and the control unit. The exchange unit manages synchronous flows input/output connections of the processor. The control unit manages the life cycle of the business component and the interactions with the runtime platform. Thus, the platform supervises all elementary processors and indirectly all business components.

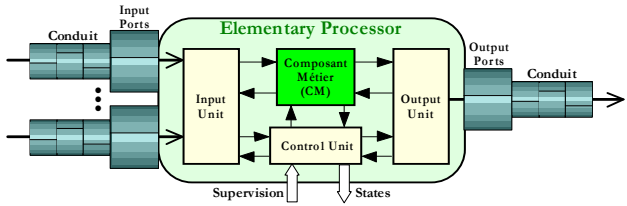

**Figure 7. Architecture of the Elementary Processor**

Thanks to the elementary processor, business components read and write synchronous slices of synchronous flows even if these slices contain several data flows. Indeed, the business component processes some flows, the others are only transferred from input unit to output unit into the processor. When processings are ended by the business component, the output unit executes synchronization policies in order to compose slices broke by input unit. This solution is the one that we propose in order to prevent the desynchronization of flows induced by processing.

Synchronous flows are transported between elementary processor by means of a connector called the conduit. Its main role is to connect software components (functional and non-functional) of applications. The conduit receives synchronous flows slices produced by components and conveys them. It is made up of two parts. The control unit implements interactions between the conduit and the platform while an exchange unit manages the input/output connections with components. The conduit is the distributed entity of our model, i.e. it can transfer synchronous flows between different sites of distributed applications. Its internal architecture is based upon the client/server model. It constitutes the solution that we propose in order to avoid the desynchronization of flows caused by network transmission.

## 5. PROTOTYPE IMPLEMENTATION

In order to validate our approach and the different models presented in this paper, we developed several prototypes. This section presents them briefly. The first one is a simulator that we used to test and validate the synchronization policies presented above. The second one is used to test the synchronization of two video flows by introducing a processing on one of them.

### 5.1 Synchronization Policies Simulator

The simulator developed allows to apply the synchronization policies on synchronous flows with both temporal constraints. It permits to provide synchronous slices of flows with only time stamps because the policies use these stamps. Moreover, the aim of the simulator is to test the constitution of the composed flows. For each flow provided, we can choose the properties of the flows. The results of the constitution of the slices of the providing composed flow are displayed on the main window of the simulator and stored in a text file. Thus, we can analyze the results after the runtime of the simulator.

In order to create synchronous flows, the simulator uses several parameters. The first one allows to define the θ value in order to differentiate the flows with both temporal constraints. The α parameter permits to express the maximal delay to wait synchronous slices in input of the operator which applies the policies. This parameter is used by the mixed and soft policies. The two last parameters allow to indicate the number of flows with hard temporal constraint and the number of flows with soft temporal constraint.

Each flow is created with three properties. The first one is an integer which represents the number of the flow. The second one gives the intra-flow synchronization relation between two successive slices. For the flows with soft temporal constraint, this is the maximal time between two slices. Indeed, for these flows this time may vary. Finally, with the last property, it is possible to introduce a maximal delay for the arrival of slices at the input of the simulator. Thus, we consider the delay between the flows.

The simulator is composed of a set of windows described on Figure 8. The main window gives the constitution of the created slices of the resulting composed flow. Each slice indicates, for each flow, the number of slices used for the constitution of the resulting slices. The slices used for this constitution are indicated by the mean of their time stamps. The time $T_{MAX}$ used to constitute a slice is indicated for each resulting slice except when the soft policy is applied because it does not use this time. The flows indicated by a capital letter are with hard temporal constraint and the flows indicated with a small letter are with soft temporal constraint. For a given slice, on flows with soft temporal constraint we can obtain the following description: {f0=}. This description means that slices are available for this flow but for the moment they do not correspond to the criteria used for the given slice. In fact, this kind of slices is available too early.

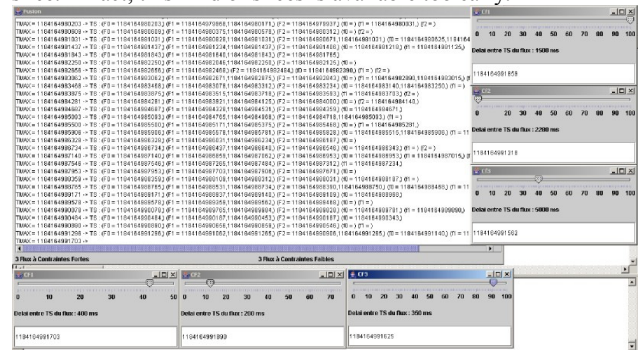

**Figure 8. Synchronization Policies Simulator**

This simulator can be used by the interested readers in order to test the synchronization policies of our model. This simulator can be downloaded at the following URL: http://www.iutbayonne.univ-pau.fr/~roose/V2/korronteaSimulator A read me file is given in order to give some explanations on how to use it.

### 5.2 Application prototype

Finally, we ended this paper by detailing a prototype developed in order to test dynamic adaptation of an application and the synchronization policies introduced here.

This prototype is distributed. It is composed of two parts. The first one implements two video capture components, one from a WebCam and the other from a file. Each of these two components produces two same video flows that are put in a composed flow in order to keep their synchronization relations. This composed flow is transmitted to the other part of the application by the mean of a distributed conduit. This part is composed of one ore more components. This composition depends on the configurations that we want to implement. We can find a displaying component which allows to display the two video flows. Possibly, we can add (before the displaying) one ore more processing components on one of the two flows. These components can be added or removed dynamically. This prototype provides six processing components: an image size reduction of the video, a negative transformation of the video, a black and white transformation of the video, an edge detection, a blured video and a skin color detection. Each component of the applications is executed in an elementary processor.

The graphic interface allows to simulate the runtime platform by adding or removing and starting and stopping the processing components. It permits equally to choose the source of the video flows (WebCam or file). This interface is shown in the Figure 9.

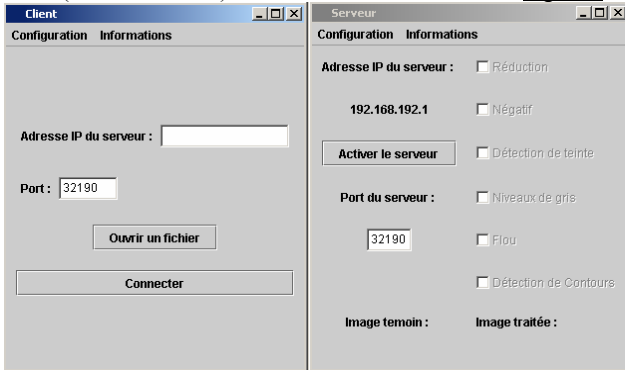

**Figure 9. Graphic Interface of the prototype**

On the left window, several fields are used to input data like IP address and the port of the machine where the application will send the flows. Two buttons allow to choose the video source. The right side is executed on another machine where the IP address is indicated. The processing components are chosen in this part of the prototype by the mean of check boxes. The displaying of the video flows are realized on the low part.

The principle of the prototype is easy to understand. The creation component produces two video flows with hard temporal constraint. Thus, the hard synchronization policy is applied. The composed flow is transmitted to the second part of the application (right side of the interface in the Figure 9). In this part of the application, it is transmitted to the displaying component after one or several processing components chosen by the user. The processing components receive the composed flow and apply their processing only on the second flow. The first one allows to know if the inter-flow synchronization relations between the two flows are kept. Moreover, the prototype allows to test the dynamic adaptation of the application.

The Figure 10 shows an example of the runtime of this prototype. In this case, we add an edge detection component and a component which applies a blur on the video. We can see on the video that they are synchronous despite the processing applied on the second flows. The Figure 11 shows the same configuration to which we added an image size reduction of the video. We can see that the two videos are synchronous too despite the three processing. We can remove processing components too. The Figure 12 shows the same example to which we removed the edge detection component. The video are synchronous one more time. The displaying component is implemented in order to consider the intra-flow synchronization relations of the two videos. To do this, we used the time stamps of the synchronous slices.

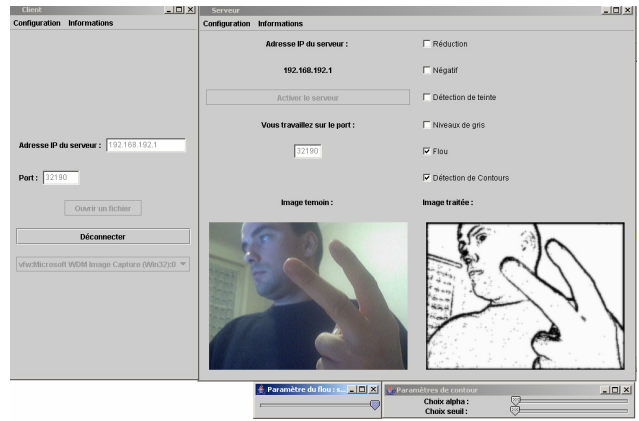

**Figure 10. Edge Detection and Blur**

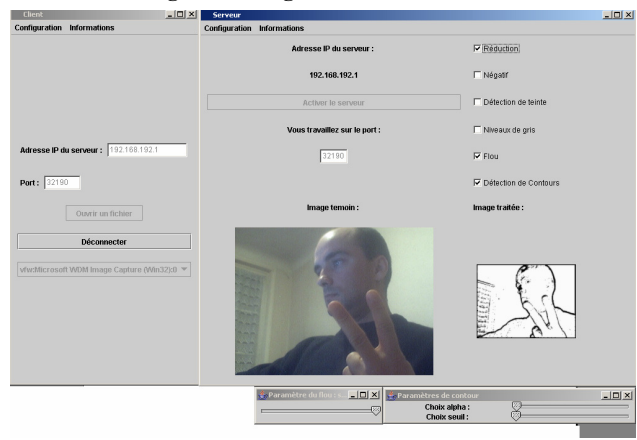

**Figure 11. Adding of an Image Size Reduction**

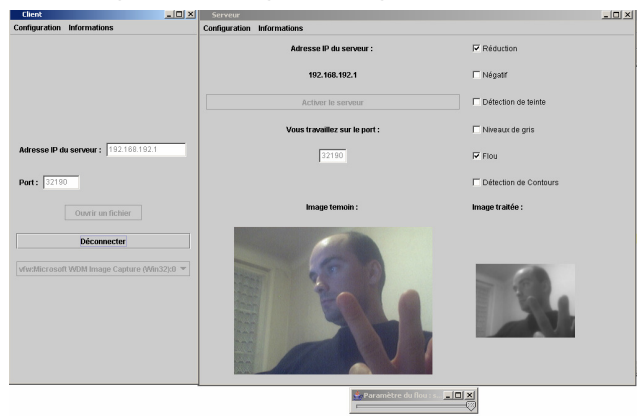

**Figure 12. Removing of the Edge Detection**

## 6. RELATED WORKS

The development of pervasive and ubiquitous computing imposes stringent requirements that the deployment of multimedia applications must consider in order to dispose of efficient implementations. These requirements impose to adapt both applications and data. For instance, many works deal with data adaptation according to the characteristics of the runtime context (transmission through networks and client peripheral capacities). These adaptations are performed by increasing or decreasing data intrinsic quality [30]. The components paradigm allows this kind of adaptation by adding components in critical locations of a distributed application. This solution is used by our platform in order to adapt data to provided QoS.

Data must be structured in order to be handled by the entities that compose an application. Some works use coding formats like the MPEG one [31]. This kind of solution allows to consider data properties but is open to criticism for many reasons. Data integration is an important criterion in multimedia applications [5]. An MPEG solution is based on the multiplexing of several media or data. Thus, this kind of solution goes against the data integration because we loose the possibility to process single data except with the implementation of complex architectures. Indeed, this solution is not efficient due to the considerable times of compression and decompression processes. Moreover, the multiplexing of several data implies to consider a global QoS for the data. Thus, the adaptation of each media by increasing or decreasing their intrinsic quality [30] is not possible. Other disadvantages of solutions oriented multiplexing are given in [32]. Consequently, we prefer to define our own data model suitable for our works. This model provides a high degree of integration. We do not describe it in this paper (not yet published) but the provided architecture uses it.

The properties of this data must be ensured in such applications. In previous paper [12], we identified two major problems of desynchronizations. The first one is due to the fact that some data synchronized with others must be processed. Processing introduces temporal delays on media and so desynchronizes the processed media in comparison with the others. The second one can occur when synchronous data are transmitted through network. Some services used to transport data on a network introduce an increase of the network load but also packet loss, delay and jitter [13]. These problems are well-known and harmful to media synchronization [14]. Some works provide models to ensure synchronous presentation of media whose characteristics are known a priori [5] (a survey is given in this paper). However, it is hard to know a priori all the characteristics of data that can be used in multimedia applications. Other works provide models to allow synchronous transmission of media through networks [33]. The RTP protocol [34] gives this possibility. These works do not consider the first problem of desynchronization. We think that synchronization models are efficient only if they are extended to the whole application in order to avoid these two problems of desynchronization. With this aim, we use a data model which allows to keep synchronization relations in real-time from the source to the sink of an application. This model introduces policies for this task.

The work presented in the following thesis [35] defines a data flow graph model to specify the flows and their temporal constraints in multimedia systems. Then, algorithms are used to translate these specifications into scheduling information. A graph representation is a structure easily adaptable [36]. We think that these solutions are interesting to perform dynamic adaptation of applications. Indeed, we can adapt an application by adapting the architecture described by the graphs. We choose a solution based on graphs to describe architecture of multimedia applications.

We choose to implement such applications by the means of software components. We think that they constitute an interesting solution for dynamic adaptation of applications. Academic [37] and commercial [38] models allow developers to design and implement applications. These models provide non-functional properties such as persistence, transaction, or security. However, there is a lack when the interest is focused on a particular domain. We think that these properties must be extended to take into consideration the characteristics of this domain. Thus, new models are provided by both industrials and researchers. For instance, the PECOS model [39] has been proposed for a specific class of embedded systems. In this way, we provide a component model whose aim is to take into consideration the specific characteristics of distributed multimedia applications.

## 7. CONCLUSION

In this paper, we presented both data and synchronization models suitable for distributed multimedia applications. From specifications given by the runtime platform, we have specified the Korrontea model used to transport data and media into applications in either local or distributed ways. We propose to use a data flow structure in order to model this data. The key idea of this research consists in exploiting the temporal dimension of the data flows that can be handled in such applications. A data flow can be built with hard or soft temporal constraint. The temporal constraint is linked to the type of the data transported into the flow. Synchronization policies that we used are based on these constraints. The Korrontea model is then used in the Osagaia component model in order to develop multimedia applications with these specifications.

The robustness of the proposed technique has been proved by the synchronization policies simulator. Another prototype allow to see that synchronization is maintained in real use case and this in spite of processings and network transfers.